# LATTICE BOLTZMANN METHOD SIMULATION OF 3-D MELTING USING DOUBLE MRT MODEL WITH INTERFACIAL TRACKING METHOD

**Zheng Li**
Department of Mechanical and Aerospace Engineering
University of Missouri
Columbia, Missouri, USA

**Mo Yang**
School of Energy and Power Engineering
University of Shanghai for Science and Technology
Shanghai, China

**Yuwen Zhang**
Department of Mechanical and Aerospace Engineering
University of Missouri
Columbia, Missouri, USA
Email: zhangyu@missouri.edu

**ABSTRACT**
Three-dimensional melting problems are investigated numerically with Lattice Boltzmann method (LBM). Regarding algorithm's accuracy and stability, Multiple-Relaxation-Time (MRT) models are employed to simplify the collision term in LBM. Temperature and velocity fields are solved with double distribution functions, respectively. 3-D melting problems are solved with double MRT models for the first time in this article. The key point for the numerical simulation of a melting problem is the methods to obtain the location of the melting front and this article uses interfacial tracking method. The interfacial tracking method combines advantages of both deforming and fixed grid approaches. The location of the melting front was obtained by calculating the energy balance at the solid-liquid interface. Various 3-D conduction controlled melting problems are solved firstly to verify the numerical method. Liquid fraction tendency and temperature distribution obtained from numerical methods agree with the analytical results well. The proposed double MRT model with interfacial tracking method is valid to solve 3-D melting problems. Different 3-D convection controlled melting problems are then solved with the proposed numerical method. Various locations of the heat surface have different melting front moving velocities, due to the natural convection effects. Rayleigh number's effects to the 3-D melting process is discussed.

## INTRODUCTION
Melting problems appear in different areas such as thermal energy storage, electronics cooling, and food processing. These problems always involve nonlinearities, strong couplings and a moving boundary [1]. Analytical, experimental and numerical methods can be used to solve melting problems. Numerical methods for melting problems are in consideration in this paper.

Bertrand et al. compared the results from different numerical methods [2]. The numerical methods to solve solid-liquid phase change problem can be divided into two groups [3]: deforming grid and fixed grid approaches [4]. Deforming grid approach uses coordinate transformation technique to transfer the solid and liquid phase geometries into the fixed regions. The high computational cost and complexity of the governing equations and boundary conditions are the main shortcomings of this method. On the other hand, the fixed grid method uses the same set of governing equations for both phases throughout the simulation. Rather than explicitly tracking the interface in the deforming grid approach, the interface is obtained from the temperature distribution in the fixed grid approach. It was reported that the fixed grid approach can reach the same accuracy as the deforming grid approach with much less computational time [5]. The enthalpy method [6, 7] and equivalent heat capacity method [8, 9] are the two major methods in the fixed grid approaches. The equivalent heat capacity method can only solve the melting problem occurring in a range of temperature; when this range is small, it is difficult to reach the converged result using this method. The enthalpy method has difficulty in temperature oscillation.

To overcome the disadvantages in equivalent heat capacity method and enthalpy method, Zhang and Chen [10] proposed an interfacial tracking method for the melting under ultrafast laser heating. It has advantages in good computational stability, high computational efficiency and applicability in melting taking place at a fixed temperature or in a range of temperature. Chen et al. and Li et al. [11, 12] applied this method to solve natural convection controlled melting in rectangular enclosures under the constant wall temperature and constant heat flux respectively. And these results are based on finite volume method (FVM). On the other hand, LBM is a promising numerical method to solve fluid flow and heat transfer problems. Different models exist in LBM to solve heat transfer problems. Li et al [13, 14] solve melting problems using FVM-LBM hybrid method and double distribution model in LBM with interfacial tracking method. All of the above applications of the interfacial tracking method are based on 2-D problems.

In this paper, double Multiple-Relaxation-Time (MRT) models [15] with interfacial tracking method is proposed to solve 3-D melting problems. Conduction melting problems are solved firstly to verify the proposed method. After it, two convection controlled 3-D melting problems are solved to discuss the Rayleigh number effects to these processes.

## NOMENCLATURE

- $e_i$    particle speed
- $f_i$    density distribution
- $F_i$    body force
- $g_i$    Energy distribution
- $M$    transform matrix for density distribution
- $m_i$    moment function for density distribution
- $N$    transform matrix for density distribution
- $n_i$    moment function for energy distribution
- Pr    Prandtl number
- $Q$    collision matrix for energy distribution
- Ra    Rayleigh number
- $S$    Interface location
- Ste    Stefan number
- $\Omega_i$    Collision term

## 1. PROBLEM STATEMENT

Physical model of melting in a cubic cavity is shown in Fig. 1. The cubic cavity with an edge length of H is filled with working fluid of PCM. The left wall (y=0) is kept at a constant temperature $T_h$, which is higher than the melting temperature $T_m$. The right wall is also kept at a constant $T_c$ that is below or equal to $T_m$. Meanwhile the top and the bottom of the enclosure are adiabatic. No slip conditions are applied to all the boundaries. The initial temperature of the system is at $T_c$. The following assumptions are made for convection controlled melting problem.

1. The PCM is pure and homogeneous.
2. The volume change due to the melting process is negligible
3. The liquid PCM is Newtonian and incompressible.
4. Boussineq approximation is applied to the liquid PCM.
5. Natural convection in the liquid PCM is laminar.

Governing equations for the liquid PCM can be found in Ref. [16].

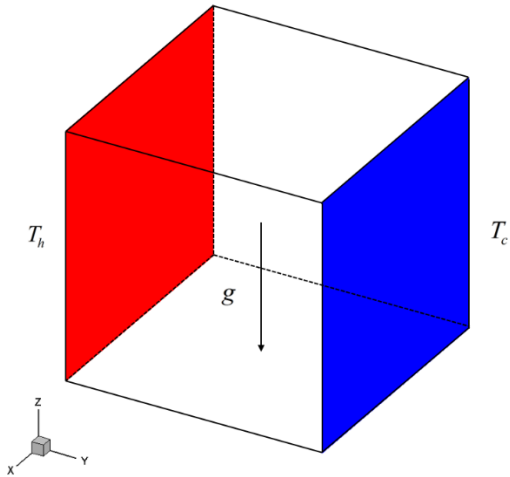

Fig. 1 Physical model

## 2. NUMERICAL METHOD

Many models exist in LBM to simulate fluid flow and heat transfer problems, and double distribution model is employed in this paper. Double MRT for 3-D natural convection problems are proposed in Ref. [16] for the first time. Fluid field and temperature field in liquid PCM in this paper are solved with D3Q19 and D3Q7 respectively.

Velocity, density and pressure are reached using density distributions in D3Q19, which are governed by Eq. (1):

$$f_i(\mathbf{r}+\mathbf{e}_i\Delta t, t+\Delta t) - f_i(\mathbf{r},t) = \Omega_i + F_i, \quad i=1,2,...19 \quad (1)$$

To satisfy the continuum and momentum conservations, the collision term in MRT is:

$$\Omega_i = -M^{-1} \cdot S \cdot \left[ m_i(\mathbf{r},t) - m_i^{eq}(\mathbf{r},t) \right], \quad i=1,2,...19 \quad (2)$$

These 19 discrete velocities are:

$$e_i = c \begin{bmatrix} 0 & 1 & -1 & 0 & 0 & 0 & 0 & 1 & -1 & 1 & -1 & 1 & -1 & 1 & -1 & 0 & 0 & 0 & 0 \\ 0 & 0 & 0 & 1 & -1 & 0 & 0 & 1 & 1 & -1 & -1 & 0 & 0 & 0 & 0 & 1 & -1 & 1 & -1 \\ 0 & 0 & 0 & 0 & 0 & 1 & -1 & 0 & 0 & 0 & 0 & 1 & 1 & -1 & -1 & 1 & 1 & -1 & -1 \end{bmatrix} \quad (3)$$

Temperature field is calculated with energy distributions, which are controlled by Eq. (4).

$$g_i(\mathbf{r}+\mathbf{u}_i\Delta t, t+\Delta t) - g_i(\mathbf{r},t) = -N^{-1} \cdot Q \cdot \left[ n_i(\mathbf{r},t) - n_i^{eq}(\mathbf{r},t) \right], \quad i=1,2,...7 \quad (4)$$

Each computational nodes have the following 7 directions:

$$u_i = c \begin{bmatrix} 0 & 1 & -1 & 0 & 0 & 0 & 0 \\ 0 & 0 & 0 & 1 & -1 & 0 & 0 \\ 0 & 0 & 0 & 0 & 0 & 1 & -1 \end{bmatrix} \quad (5)$$

All the parameters settings in LBM are the same as Ref. [16]. This double MRT model is then advanced to solve 3-D melting problem.

The key to solve melting problem is how to obtain the interface location, and interfacial tracking method is employed in this paper. Its 2-D applications are discussed in Refs. [10-14]. This method is applied to solve a 3-D problem for the first time in this paper. Different from the 2-D cases, melting front moves at the following velocity.

$$y = s, \left[ 1 + \left(\frac{\partial s}{\partial x}\right)^2 + \left(\frac{\partial s}{\partial z}\right)^2 \right] \left[ k_s \frac{\partial T_s}{\partial y} - k_l \frac{\partial T_l}{\partial y} \right] = \rho_l h_{sl} \frac{\partial s}{\partial t} \quad (6)$$

Other settings are the same as that in 2-D cases in Refs. [10-14].

## 3. RESULTS AND DISCUSSION

Various conduction and convection melting problems in 3-D are solved using LBM with interfacial tracking method. The conduction melting results are compared with analytical ones to verify the numerical method and convection melting problems in a cubic cavity are solve to discuss the 3-D characters in these processes.

### 3.1 CONDUCTION MELTING PROBLEM

For the case the natural convection is negligible, the melting front moves at a same velocity at any height. Then this problem can be simplified to a 1-D problem governed by conduction. There is an analytical solution for this problem when subcooling is 0 [2].

The cases for Stefan number equaling 0.1, 0.5 and 1 are solved for validation. To compare with analytical results, $Fo$ is used to express the non-dimensional time. Figure 2 shows the liquid fraction comparison between numerical and analytical results and they agree with each other very well for various cases. The melting process carries on faster with the *Ste* increasing.

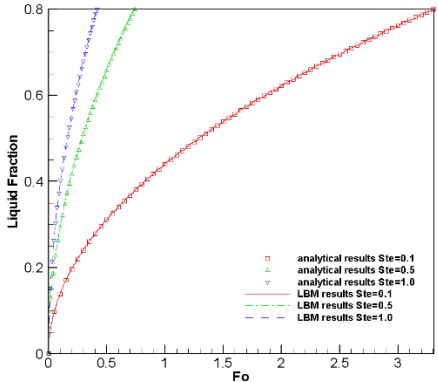

Fig. 2 Liquid fraction comparison

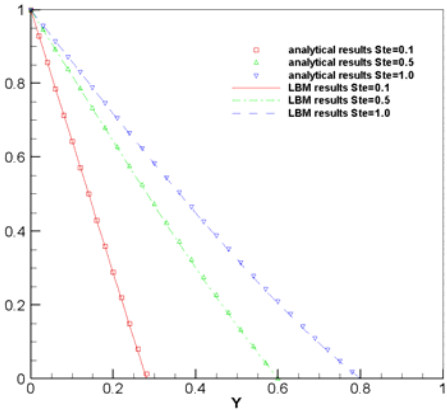

Fig. 3 Temperature distribution comparison

Figure 3 compares the temperature distribution between numerical and analytical results. The case with higher Stefan number has shorter time to reach the same interfacial location. The temperature distribution is closer to a straight line with lower Stefan number. And it is difficult to distinguish the LBM results from the analytical result for all three cases.

The numerical results agree with analytical ones well for all three cases. The LBM with interfacial tracking method is reliable for 3-D conduction controlled melting problem. Double LBM-MRT model is valid to solve various 3-Dconvection problems shown in Ref. [16]. It's reasonable to believe double LBM-MRT model with interfacial tracking method is valid to solve 3-D convection controlled melting problems.

### 3.2 CONVECTION MELTING PROBLEM

No subcooling is in consideration and three melting cases are solved using LBM with interfacial tracking method. The non-dimensional parameters in these cases are listed in Table 1.

Table 1 Two convection melting cases

|  | Pr | Ra | Ste |
|---|---|---|---|
| Case 1 | 0.02 | 25000 | 0.1 |
| Case 2 | 0.02 | 10000 | 0.1 |

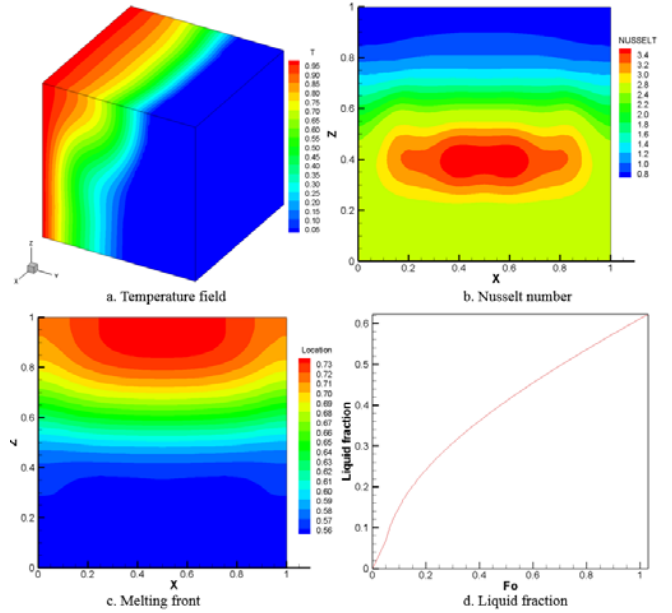

Fig. 4 Case 1 results *Fo* = 1.00

For the convection melting cases, Fourier number *Fo* is employed as the nondimensional time, which is defined as:

$$Fo = t \cdot \sqrt{3}c_S / H \cdot Ma / \sqrt{3Ra \cdot Pr} \qquad (7)$$

where $c_s$ is the sound speed, Mach number can be any number lower than 0.3, which will not affect the final results. Figure 4 shows case 1 results for *Fo* = 1.00. Temperature in liquid PCM grows with the height Z increasing as shown in Fig. 4 (a). It indicates that convection have controlled the melting process. Nusselt number distribution on the heat surface and melting front location are shown in Figs. 4 (b) and (c), which show clear three-dimensional characters. For X=0.5, Nusselt numbers are greater and melting fronts also moves faster than the other regions. Non-slip boundary conditions are applied to all the boundaries. The side walls (X=0 and X=1) slow the fluid flow closed to them. Consequently, the convection effect is decreased. Liquid fraction tendency is included in Fig. 4 (d). Comparing with the 2-D results in Refs. [12-14], this melting process is governed by conduction at the beginning and convection controlled this process later. Quasi-steady melting process is reached when the liquid fraction is linear to *Fo*.

Figure 5 shows the case 2 results for *Fo* = 1.57. These two cases are the same time regarding the *Fo* definition. Comparing with the case 1, only *Ra* is decreased to 10000. Convection also has controlled this melting process and the liquid PCM temperature grows with the height increasing. From the heat surface Nusselt numbers are lower than that in case 1, which indicates a weaker convection effect. The melting fronts at the same height don't change as much as that in case 1, which means the side wall effects are not that valid. Liquid fraction tendency is similar to that in case1 and melting process is slower than last case.

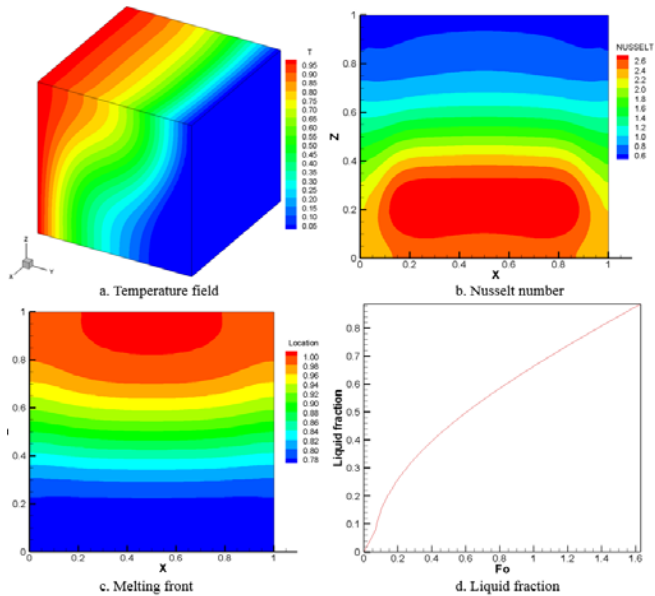

Fig. 5 Case 2 results *Fo* = 1.57

## 4. CONCLUSION

Double MRT model with interfacial tracking method is proposed to solve 3-D melting problems. Numerical results in three conduction melting problems agree with the analytical results well. The double LBM-MRT model with interfacial tracking method is valid to solve 3-D melting problems. Two convection melting problems in a cubic cavity are also solved. Convection effects in 3-D melting problems are discussed. With a lower Rayleigh number, the convection effects are weaker; side wall effects are smaller; melting process carries on slower.

## ACKNOWLEDGMENTS

Support for this work by Chinese National Natural Science Foundations under Grants 51129602 and 51276118 are gratefully acknowledged.

## REFERENCES

[1] Hannoun N., Alexiades V. and Mai T., 2003, "Resolving the Controversy over Tin and Gallium Melting in a Rectangular Cavity Heated from the Side," *Numer. Heat Tr. B-Fund.,* Vol. 44(3), pp. 253-275.
[2] Bertrand O., Binet B., Combeau H., Couturier S., Delannoy Y., Gobin D., Lacroix M., Quere P., Medale M., Mencinger J., Sadat H. and Vieira G., 1999, "Melting Driven by Natural Convection A Comparison Exercise: First Results," *Int. J. Therm. Sci.,* Vol. 38(1), pp. 5-26.
[3] Faghri A. and Zhang Y., 2006, Transport Phenomena in Multiphase System, Elsevier, Burlington, MA.
[4] Voller V., 1997, "An Overview of Numerical Methods for Solving Phase Change Problems," Minkowycz W. and Sparrow E., eds., *Advances in Numerical Heat Transfer,* 1, Taylor &Francis, Basingstoke.
[5] Ho C. and Viskanta R., 1984, "Heat Transfer during Melting from an Isothermal Vertical Wall," *J. Heat Transf.,* Vol. 106, pp. 12-19.
[6] Shamsundar N. and Sparrow E., 1975, "Analysis of Multidimensional Conduction Phase Change via the Enthalpy Model," *J. Heat Transf.,* Vol. 97, pp. 333-340.
[7] Voller V. and Prakash C., 1987, "A Fixed Grid Numerical Modeling Methodology for Convection-Diffusion Mushy Region Phase-Change Problems," *Int. J. Heat Mass Tran.* Vol. 30, pp. 1709-1719.
[8] Morgan K., 1981, "A Numerical Analysis of Freezing and Melting with Convection," *Comput. Method Appl. M.,* Vol. 28, pp. 275-284.
[9] Hsiao J., 1985, "Analyses of Heat Transfer with Melting and Solidification," *Numer. Heat Transfer,* Vol. 8, pp. 653-666.
[10] Zhang Y. and Chen J., 2008, "An Interfacial Tracking Method for Ultrashort Pulse Laser Melting and Resolidification of a Thin Metal Film," *J. Heat Transf.,* Vol. 130, p. 062401.
[11] Chen Q., Zhang Y. and Yang M., 2011, "An Interfacial Tracking Model for Convection-controlled Melting Problem," *Numer. Heat Tr. B-Fund.,* Vol. 59(3), pp. 209-225.
[12] Li Z., Yang M., Chen Q. and Zhang Y., 2013, "Numerical Solution of Melting in a Discretely Heated Enclosure Using an Interfacial Tracking Method," *Numer. Heat Tr. A-Appl.,* Vol. 64(11), pp. 841-857.
[13] Li Z., Yang M., and Zhang Y., 2014, "A Hybrid Lattice Boltzmann and Finite Volume Method for Melting with Natural Convection," *Numer. Heat Tr. B-Fund.,* Vol. 66 (4), pp. 307-325.
[14] Li Z., Yang M. and Zhang Y., 2015, Numerical Simulation of Melting Problems Using the Lattice Boltzmann Method with the Interfacial Tracking Method, *Numer. Heat Tr.A-Appl.,* Vol. 68 (11), pp. 1175-1197.
[15] Wang J., Wang D., Lallemand P. and Luo L., 2013, "Lattice Boltzmann Simulations of Thermal Convective Flows in Two Dimensions," *Comput. Math. Appl.,* Vol. 65, pp. 262-286.
[16] Li Z., Yang M. and Zhang Y., 2016, "Lattice Boltzmann Method Simulation of 3-D Natural Convection with Double MRT Model," *Int. J. Heat Mass Tran.* Vol. 94, pp. 222-238.